# Modelling Soil Water Content in a Tomato Field: Proximal Gamma Ray Spectroscopy and Soil–Crop System Models

**Virginia Strati** [1,2,*], **Matteo Albéri** [1,3], **Stefano Anconelli** [4], **Marica Baldoncini** [1,3], **Marco Bittelli** [5], **Carlo Bottardi** [1,2], **Enrico Chiarelli** [1,3], **Barbara Fabbri** [1], **Vincenzo Guidi** [1], **Kassandra Giulia Cristina Raptis** [1,3], **Domenico Solimando** [4], **Fausto Tomei** [6], **Giulia Villani** [5], and **Fabio Mantovani**[1,2]

[1] Department of Physics and Earth Sciences, University of Ferrara, Via Saragat 1, 44121 Ferrara, Italy; alberi@fe.infn.it (M.A.); baldoncini@fe.infn.it (M.Ba.); barbara.fabbri@unife.it (B.F.); guidi@fe.infn.it (V.G.)
[2] INFN, Ferrara Section, Via Saragat 1, 44121 Ferrara, Italy; bottardi@fe.infn.it (C.B.); mantovani@fe.infn.it (F.M.)
[3] INFN, Legnaro National Laboratories, Viale dell'Università 2, 35020 Padua, Italy; enrico.chiarelli@student.unife.it (E.C.); kassandra.raptis@lnl.infn.it (K.G.C.R.)
[4] Consorzio Bonifica CER, Via Masi 8, 40137, Bologna, Italy; anconelli@consorziocer.it (S.A.); solimando@consorziocer.it (D.S.)
[5] Department of Agricultural and Food Sciences (DISTAL), University of Bologna, Viale Fanin 44, 40127 Bologna, Italy; marco.bittelli@unibo.it (M.Bi.); gvillani@arpa.emr.it (G.V.)
[6] Servizio Idro-Meteo-Clima di Bologna Agenzia Regionale Prevenzione, Ambiente ed Energia, Via Po 5, 40139 Bologna, Italy; ftomei@arpae.it (F.T.)
* Correspondence: strati@fe.infn.it; Tel.: +39-3489356603

**Abstract:** Proximal soil sensors are taking hold in the understanding of soil hydrogeological processes involved in precision agriculture. In this context, permanently installed gamma ray spectroscopy stations represent one of the best space–time trade off methods at field scale. This study proved the feasibility and reliability of soil water content monitoring through a seven-month continuous acquisition of terrestrial gamma radiation in a tomato test field. By employing a 1 L sodium iodide detector placed at a height of 2.25 m, we investigated the gamma signal coming from an area having a ~25 m radius and from a depth of approximately 30 cm. Experimental values, inferred after a calibration measurement and corrected for the presence of biomass, were corroborated with gravimetric data acquired under different soil moisture conditions, giving an average absolute discrepancy of about 2%. A quantitative comparison was carried out with data simulated by AquaCrop, CRITeRIA, and IRRINET soil–crop system models. The different goodness of fit obtained in bare soil condition and during the vegetated period highlighted that CRITeRIA showed the best agreement with the experimental data over the entire data-taking period while, in presence of the tomato crop, IRRINET provided the best results.

**Keywords:** soil water content; proximal gamma ray spectroscopy; soil–crop system models; real-time soil water content monitoring; non-destructive methods; CRITeRIA; AquaCrop; IRRINET; tomato crop.

## 1. Introduction

In the context of current global warming and uncertainty about future climate conditions, variation in rainfall amounts and dry spells frequency are expected to have negative impacts on the vegetation water availability and consequently on crop yields and water productivity [1,2]. Irrigation water, the primary input for agriculture development, is about two-thirds of the total fresh water assigned to human uses [3,4]. In areas of water scarcity, such as those of the Mediterranean basin, farmers are encouraged to adopt management strategies aimed at reducing water wastes. Significant improvements typically come from advanced technological systems aimed at optimizing water uptake both in terms of amounts of water required for the specific crop and of irrigation frequency [5]. Soil–crop systems models, designed for practical applications and decision support, are useful tools for providing a real-time estimate of the amount of water available to the crop [6]. Nevertheless, the high temporal variation of soil properties, weather, and environmental conditions makes the modelling of soil water dynamics complex and problematic if not assisted with the running of expensive and time-consuming field monitoring of inter-temporal and site-specific parameters.

The continuous determination of soil moisture dynamics at field scale with non-invasive and contactless measurement techniques is at the time the most promising challenge for optimizing agricultural management and in particular for a sustainable use of water [7]. Among the proximal sensing techniques, gamma ray spectroscopy is recognized as one of the best space–time trade off methods [8,9]. Thanks to the permanent installation of measurement stations, soil water content can be estimated in real time on the basis of temporal changes in gamma ray intensity to which soil moisture is negatively correlated. Although the possibility of performing non-destructive and fast measurements has been investigated in the past decades [10–13], recent developments in cyber-physical systems boosted by investments in Industry 4.0 sectors and supported by the cloud technology are opening new perspectives in the field. In this scenario, the gamma ray spectroscopy method excellently matches the current needs for high accuracy nonstop soil water content monitoring representing a joining link between punctual and satellite fields of view. Indeed, continuous time series of water content at intermediate spatial scales are usually lacking with traditional techniques. From one side, time domain reflectometry [14] and remote sensing such as microwave scatterometers [15] provide the temporal evolution of soil water budget at punctual or catchment scale. On the other side, ground penetrating radar [16] and electrical resistivity tomography [17] act at the field spatial scale with the limitation of providing data referred only to the time of the survey.

The aim of this study is to investigate the potentialities of proximal gamma ray spectroscopy for a real-time and continuous monitoring of soil water content in the framework of an ad hoc experiment. A permanent measurement station was installed in a tomato field in the Emilia-Romagna region (Italy) and remotely controlled during the entire seven months data-taking period. Tomatoes are among the most water-intensive vegetable crops and this region represents the largest hub for its production in Italy [18]. Limited irrigation capacity, as well as water volumes excesses, can negatively influence the development of plants and fruits and provokes different kinds of diseases. Experimental daily values of soil water content, estimated on the basis of gamma-ray measurements, were compared with the results obtained with three different soil–crop system models: (i) CRITeRIA, a physically-based numerical model for soil water balance and crop systems, developed by the Regional Environmental Protection Agency (ARPA) of the Emilia-Romagna region [19]; (ii) AquaCrop, the Food and Agriculture Organization (FAO) crop model simulating the yield response to water [20]; and (iii) IRRINET, a system for irrigation scheduling developed by the Emiliano-Romagnolo Canal (CER) irrigation consortium [21].

**2. Materials and Methods**

**2.1. Experimental Site**

The proximal gamma ray sensing experiment was conducted in the period 4 April–2 November 2017 in a 40 × 108 m tomato test field of the Acqua Campus (44.57° N, 11.53° E; 16 m above sea level), a research Centre close to Bologna of the CER irrigation district in Emilia-Romagna [22] (Figure 1). The climate of the experimental area is classified according to the Koppen–Geiger classification as Cfa [23], i.e. a temperate climate, without dry seasons and with hot summers, with a mean temperature of 14.0 °C and mean annual precipitation of about 700 mm.

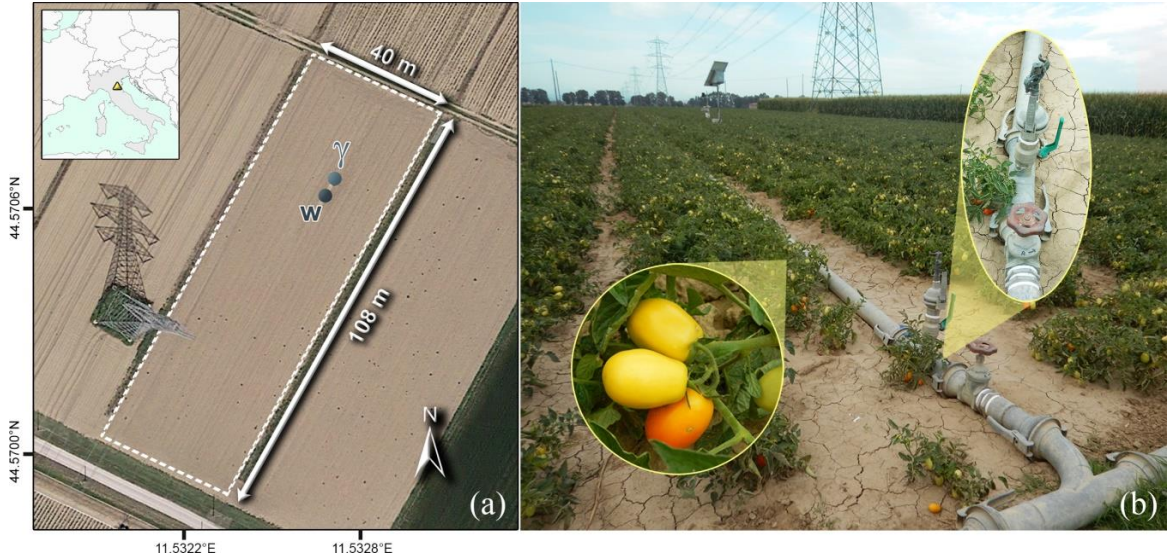

**Figure 1.** We report the geographic location of the experimental site of the Acqua Campus of CER (Emilia-Romagna, Italy), the dimensions of the tomato test field and the positions of the gamma ($\gamma$) and agro-meteorological (w) stations (panel **a**). The tomato crop with focuses on a tomato plant and on a sprinkler are shown (panel **b**).

The main physical and hydraulic parameters of the soil, characterized by a loamy texture, are reported in Table 1. The analysis by the sieving and the hydrometer method followed the standard procedure described in [24] and was performed with the aim of measuring the particle size distribution (PSD) function and, therefore, the percentage of silt, clay, and sand (Table 1). The soil classification was based on the United States Department of Agriculture (USDA) method. The PSD was used to derive the soil water retention and hydraulic conductivity functions.

**Table 1.** Physical and hydraulic parameters of the experimental site soil for the depth horizon [0–30] cm. Sand, silt, and clay percentage as well as bulk density and organic matter were determined from direct measurements. The wilting point ($\theta_{WP}$), field capacity ($\theta_{FC}$), saturation value ($\theta_s$), and saturated hydraulic conductivity ($K_s$) are inferred on the basis of the methods discussed in Section 2.1.

| Parameter | Value |
|---|---|
| Sand (%) | 45 |
| Silt (%) | 40 |
| Clay (%) | 15 |
| Soil textural class | Loam |
| Soil bulk density (kg/m$^3$) | 1345 |
| Organic matter (%) | 1.26 |
| Wilting Point ($\theta_{WP}$) (m$^3$/m$^3$) | 0.09 |
| Field Capacity ($\theta_{FC}$) (m$^3$/m$^3$) | 0.32 |
| Saturation ($\theta_s$) (m$^3$/m$^3$) | 0.48 |
| $K_s$ (cm/day) | 23 |

After having experimentally obtained the PSD, we employed a mono-modal log-normal distribution, where the particle diameter is replaced by its natural logarithm. Details about the methodology are provided in [25]. The $\mu$ = 22.07 μm mean and a $\sigma$ = 2.81 μm standard deviation show that the distribution is within the silt range, when using the USDA textural classifications limits [25]. The obtained PSD parameters were adopted for the retrieving the parameters of the Campbell's equation [26]:

$$\psi_e = 0.61\ln\mu - 3.9 \tag{1}$$

$$b = 8.25 - 1.26\ln\mu \tag{2}$$

where $\psi_e$ is the air entry potential (J/kg), $b$ is the slope parameter in the Campbell's water retention curve equation [27]. The saturated water content was obtained from the relationship $\theta_s = 1 - \dfrac{\rho_b}{\rho_s}$, where $\rho_b$ is the measured bulk density (Table 1), and $\rho_s$ is the particle density (a value of 2600 kg/m³ was assumed). The Campbell's water retention curve (Figure 2) for the soil water content $\theta$ as function of the $\psi$ potential was then obtained as:

$$\theta(\psi) = \begin{cases} \theta_s \left(\dfrac{\psi_m}{\psi_e}\right)^{-\frac{1}{b}}, & \text{if } \psi_m \geq \psi_e \\ \theta_s, & \text{if } \psi_m < \psi_e \end{cases} \tag{3}$$

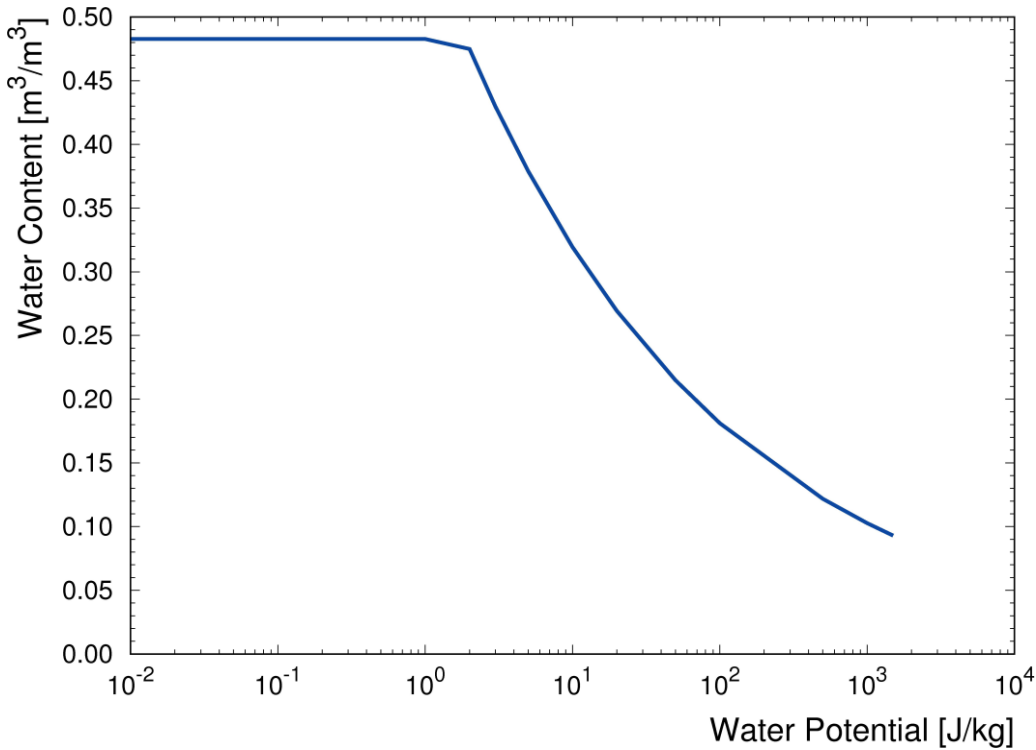

**Figure 2.** Campbell's water retention curve obtained on the basis of the mean particle size determined after the parameterization of the PSD.

The soil hydraulic properties in terms of wilting point ($\theta_{WP}$), field capacity ($\theta_{FC}$), and saturated hydraulic conductivity ($K_s$) (Table 1) were then inferred from the retention curve of Figure 2.

Tomato plants were transplanted on the 23 May in coupled rows: the plants geometry was such that the distance between two couples was 1.5 m, the distance between two rows of the same couple 0.45 m and the distance between two tomato plants of the same row 0.38 m. On the basis of this configuration, and assuming a homogeneous plant distribution on the experimental site area, a 3.5 plants/m² planting density was estimated. The anthesis of the plants took place approximately on the 9 June, the berries maximum maturity occurred on the 30 August and the harvesting on the 14 September (Figure 3a). The crop irrigation, performed via sprinklers (Figure 1), was scheduled following the criteria provided by the decision support tool of IRRINET (see Section 2.3.3).

## 2.2. Experimental Setup

The experimental setup comprised a commercial agro-meteorological station (MeteoSense 2.0, Netsens) and a gamma ray spectroscopy station specifically designed and built for the experiment (Figure 4).

Measured weather data included air temperature, relative air humidity (RH), wind direction and speed, precipitation, and short wave incoming radiation (SWIR), the latter measured with a silicon-photodiode pyranometer. We report for the entire data-taking period the temporal profiles of daily values recorded for the minimum ($T_{min}$) and maximum temperatures ($T_{max}$) (respectively ranging in the $T_{min}$ = (1.3–22.7) °C and $T_{max}$ = (13.5–39.3) °C intervals) (Figure 3a), the SWIR (ranging from 34.7 to 257.3 W/m$^2$) (Figure 3b), the RH (ranging from 44.3% to 95.2%), the rainfall amount (which reached a maximum of 56.2 mm), the irrigation water (which reached a maximum of 35 mm) (Figure 3c). The reference evapotranspiration (ET0) (Figure 3b) was calculated through the Hargreaves equation [28] by adopting the weather parameters recorded during the data-taking by the agro-meteorological station.

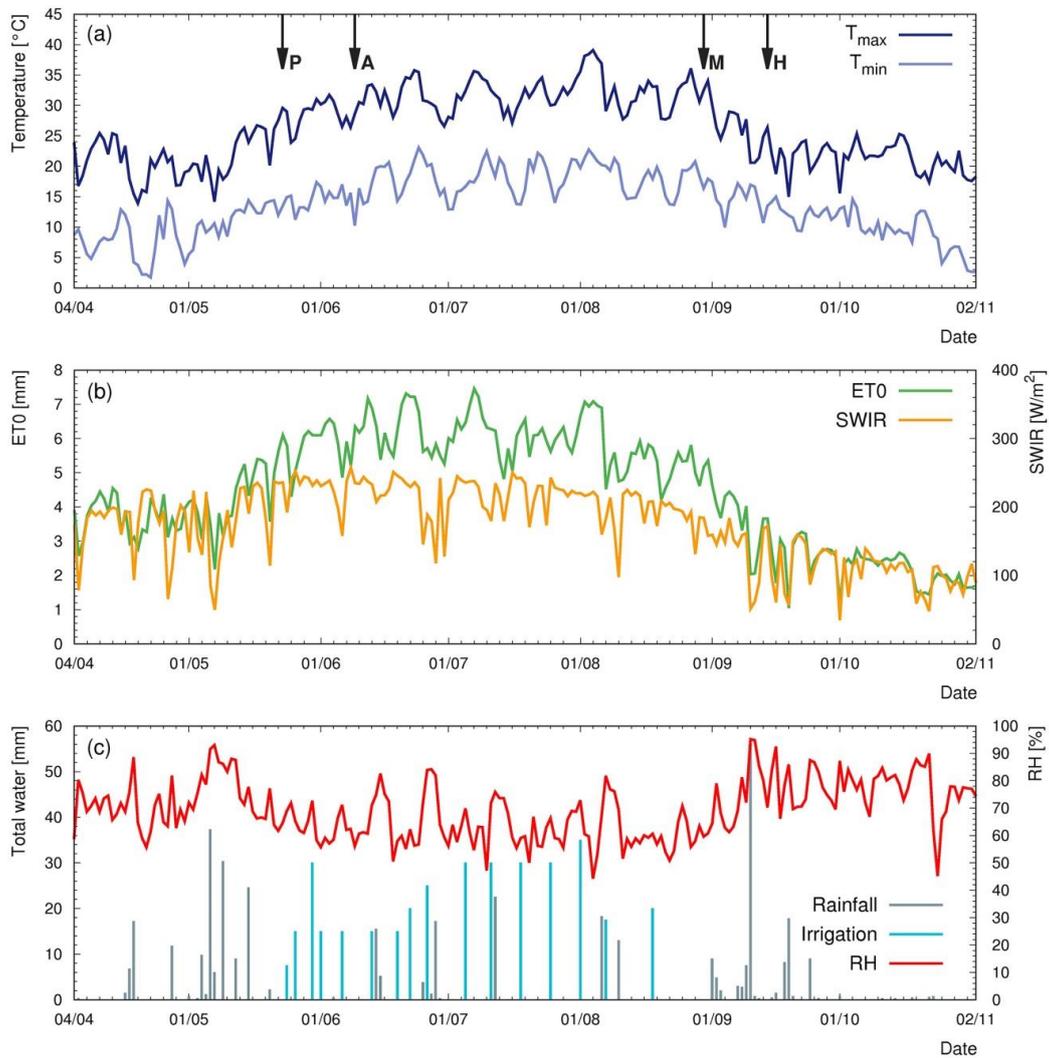

**Figure 3.** Weather parameters recorded in the 4 April–2 November 2017 data-taking period by the agro-meteorological station installed at the experimental site: maximum ($T_{max}$) and minimum ($T_{min}$) temperature (panel **a**), short wave incoming radiation (SWIR) (panel **b**) and relative air humidity (RH), rainfall and irrigation water amount (panel **c**). The reference evapotranspiration (ET0) (panel **b**) is calculated on the basis of Hargreaves equation [28]. The arrows in panel **a**) indicate the four major crop maturity phases, i.e., planting (P, 23 May), anthesis (A, 9 June), maturity (M, 30 August), and harvesting (H, 14 September).

A permanent gamma station, specifically designed and built for the experiment, with a 1 L sodium iodide (NaI) crystal was placed inside a steel box mounted on top of a 2.25 m high steel pole [29]. The gamma spectrometer was coupled to a photomultiplier tube base which output is processed by a digital multi-channel analyzer (MCA, CAEN γstream); the whole system was powered by a solar panel (Figure 4). The detector measured the photon radiation produced in the decays of natural occurring radionuclides ($^{40}K$, $^{238}U$, and $^{232}Th$) and recorded a gamma spectrum, i.e., a histogram representing the energy distribution of photons emitted by the source. Since each gamma decay has a specific emission energy, a gamma spectrum is characterized by the presence of distinctive structures (photopeaks) which allow for the identification and quantification of $^{40}K$, $^{238}U$, and $^{232}Th$ abundances in the soil. The integrated number of events inside the energy ranges associated to the main photopeaks [30] were used to determine the counts per second (cps), associated to $^{40}K$, $^{238}U$, and $^{232}Th$ activities in the soil. The statistical uncertainty on the measured cps for a gamma spectrum with a temporal length of one hour is typically lower than 1%.

The lateral and vertical horizons of proximal gamma-ray spectroscopy can be defined on the basis of the probability law that governs photon survival in traversing a given material, as reported in [31]. As photon propagation is ruled by the density of traversed materials, a gamma spectrometer is sensitive to about 25 m far in the horizontal direction (Figure 4) and approximately 30 cm deep in the soil (Figure 5).

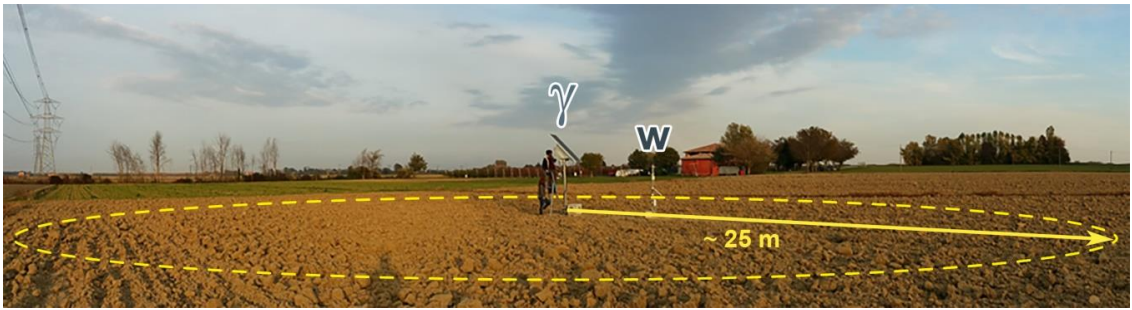

**Figure 4.** Gamma (γ) and agro-meteorological (w) stations installed at the experimental site. About 95% of the signal received by the gamma spectrometer located at 2.25 m above the ground is produced within a ~25 m radial distance [31].

Weather data were recorded by the agro-meteorological station with an average frequency of about five minutes, while the gamma station provided a list mode output, i.e., a continuous logging of individual photons arrival time and energy. Therefore, a dedicated software was developed to group and synchronize all the acquired data in a single time-referenced dataset. As both stations were equipped with a GPRS connection, it was possible to remotely preprocess the data in real time. The resulting dataset had a temporal resolution of 15 minutes and merged 10 data fields from the agro-meteorological station temporally aligned with 34 data fields from the processed energy spectra. Both stations were operative during the entire data taking period, for a 94.8% overlapping duty cycle and a 260 GB global amount of uncompressed data, the major part being raw gamma spectra.

**2.3 Soil water content with soil crop system models**

The daily values of soil water content simulated with CRITeRIA, AquaCrop, and IRRINET soil–crop system models were compared to the experimental daily values inferred from proximal gamma ray spectroscopy. For the simulation we used the soil properties reported in Table 1, daily temperatures and precipitation amounts measured by the agro-meteorological station, scheduled irrigations, growth stages of the tomato crop and reference evapotranspiration (ET0) calculated on the basis of the Hargreaves method [28] (Figure 3). The CRITeRIA and AquaCrop values, referred to the entire data-taking period (4 April–2 November 2017), were obtained by running the models with a six month spin-up which was performed by adopting meteorological data published at [32] in order to adjust for initial conditions [25]. As the IRRINET simulation is strictly bounded to the crop development, soil water content data are referred to the sole tomato crop season (23 May-14 September 2017).

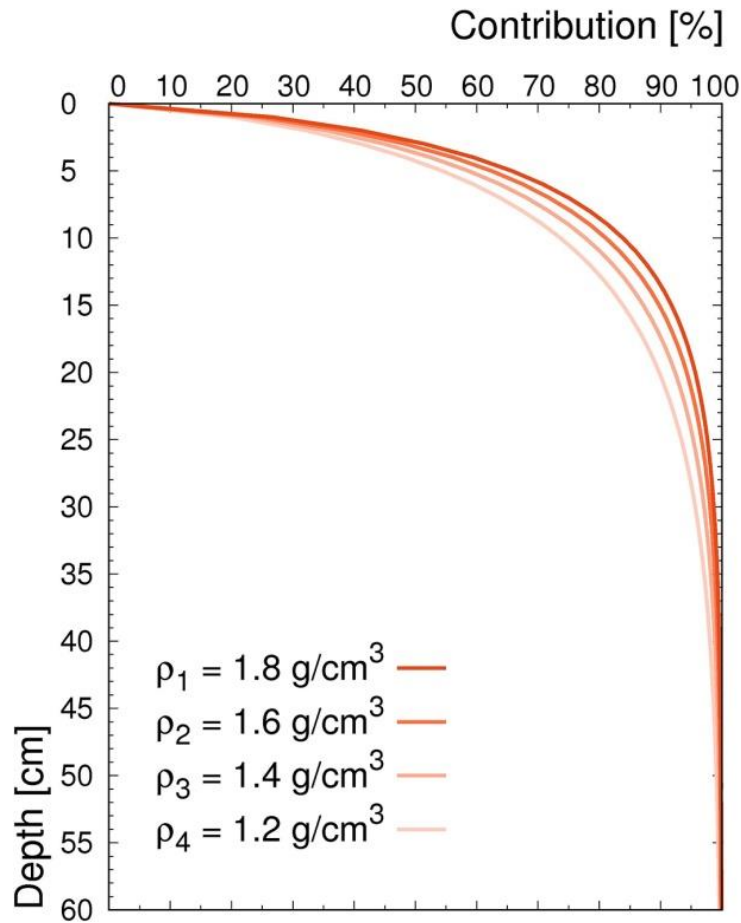

**Figure 5.** Percentage contribution to the $^{40}$K ground level unscattered photon flux as function of soil depth. The contribution is computed, adopting the theoretical background reported in [31], for four soil density values: as the density increases from 1.2 g/cm$^3$ to 1.8 g/cm$^3$, the depth corresponding to 95% of the signal decreases from 28 cm to 19 cm.

**2.3.1. CRITeRIA**

CRITeRIA is a suite of soil water balance and crop modelling systems developed by the Regional Environmental Protection Agency (ARPA) of the Emilia-Romagna region (Italy). Here we used CRITeRIA-1D, which is a one-dimensional model simulating the soil water balance, the nitrogen balance and crop development. The CRITeRIA-1D model simulates soil water movement by using a numerical solution of Richards' equation as described in [19]. The model implements many herbaceous crops and fruit trees and requires as input daily weather data (temperature and precipitation), soil texture and hydraulic properties, bulk density and crop management information. For this study, water retention and hydraulic conductivity curves were determined on the basis of PSD experimental data (see Section 2.1).

CRITeRIA implements a crop growth model based on the day degree sum. The relevant variables are the development of the leaf system (expressed by the leaf area index parameter) for the epigeal part and a root growth model for increase and spatial distribution of the root system. As the tomato season was characterized by anomalously high temperatures, parameters regulating the leaf area index increase and decrease were properly calibrated to follow the effective tomato plant growth at the experimental site.

### 2.3.2. AquaCrop

AquaCrop is a crop water productivity model implemented by the Land and Water Division of FAO [20]. In the perspective of being applied to diverse worldwide agricultural systems characterized by large crop and soil variability and by the unavailability of extended number of site specific input variables, AquaCrop is structured as an almost ready to use simulation model, requiring a low number of explicit parameters.

The model simulates the water balance and yield response to water of herbaceous crops and it is typically used in situations where water is a key limiting factor in crop production [20]. AquaCrop develops a structure (sub-model components) that includes: (i) the soil, with its water balance; (ii) the crop, with its development, growth, yield and management; (iii) the atmosphere, with its thermal regime, rainfall, evaporative demand, and carbon dioxide concentration. The calculation procedure is grounded on basic and complex biophysical processes to guarantee an accurate simulation of the response of the crop in the plant–soil system. The computation of the soil water transport is based on a tipping-bucket conceptual model employing soil hydraulic properties in terms of $\theta_{FC}$ and $\theta_{WP}$.

### 2.3.3. IRRINET

IRRINET is one of the tools provided to farmers in the framework of the Emilia-Romagna Action Plan for Rural Development 2007–2013. It is a model for irrigation management backed by the results of more than 50 years of research on plant/water relation and sustainable irrigation management [21]. IRRINET was developed with the aim of progressively reducing water use for irrigation without harming farmers' income, therefore saving water and optimizing water productivity. The freely available service is web and GIS based and provides decisional criteria for irrigations for a large number of water demanding crops. The model can be employed by using several data sources as meteorological and soil data from local services and crop parameters as defined by CER, including application of the most effective crop tailored irrigation strategy.

Since 2009, IRRINET implements economic calculation of the irrigation profitability assessing the economic benefit related to the next irrigation. Users are provided with optimal irrigation volume and interval, via web or mobile phone text message.

The crop water balance is calculated at field scale with a daily frequency and referred to the crop characteristics, simulated, or input by the farmer. The model structure deals with the soil–plant–atmosphere continuum and includes (i) the soil water balance with capillary rise [33], infiltration rate and run off [34]; (ii) the development of plant with crop coefficients [35]; (iii) the atmosphere (thermal regime, rainfall, and evaporative demand); and (iv) the irrigation system adopted. For sprinkler systems, IRRINET guides the irrigation in order to maintain soil moisture between 30% and 70% of the available water ($\theta_{FC}$-$\theta_{WP}$). Mean soil water content is estimated in a progressively deeper soil range which linearly follows the growth of the root system from an initial depth of 15 cm (at the time of transplanting) up to 65 cm.

## 3. Results

The soil water content was determined on the basis of proximal gamma-ray spectroscopy for the entire seven month data-taking period. The basic principle for the determination of soil water content using natural terrestrial gamma radiation recorded by a permanent measurement station consists in the quantification of relative differences between gamma signals measured under different moisture level conditions. While $^{208}$Tl ($^{232}$Th decay chain) and $^{40}$K are distributed solely in the soil, gamma radiation originated by the decay of $^{214}$Bi ($^{238}$U decay chain) comes both from $^{214}$Bi in the soil and from $^{214}$Bi in the atmosphere produced by the decay of $^{222}$Rn gas exhaled from rocks and soils. For this reason, we chose $^{40}$K as natural gamma emitter for soil water content assessment purposes, given also the highest net counting statistics in its photopeak.

Since the water mass attenuation coefficient is significantly higher than those of typical minerals commonly present in the soil, a gamma spectroscopy measurement is extremely sensitive to different soil moisture levels. The gamma photon flux at the soil–air interface and, consequently, the measured gamma signal are inversely proportional to soil water content: by simultaneously performing an independent gravimetric calibration measurement and a radiometric acquisition, it is possible to evaluate the gravimetric soil water content $w_\gamma$ (kg/kg) at a specific time $t$ as [29,36]:

$$w_{\gamma K}(t) = \frac{S_K^{Cal}}{S_K(t)} \times \left[\Omega + w^{Cal}\right] - \Omega \qquad (4)$$

where $w^{Cal}$ is the gravimetric water content at calibration time, $S_K^{Cal}$ and $S_K(t)$ are the gamma signals in cps recorded in the $^{40}$K energy window respectively at calibration time and at time $t$.

The calibration measurement $w^{Cal}$ = (0.163 ± 0.008) kg/kg is the mean gravimetric water content estimated after a gravimetric campaign performed in the experimental field in bare soil condition in the 0–30 cm depth range at 16 planar sampling points homogeneously distributed within 15 m from the gamma station. The gravimetric measurements of the calibration process are described in detail in [36]. The gamma calibration measurement $S_K^{Cal}$ refers to a 2 h late morning data acquisition (10.00 a.m.–12.00 a.m.) concomitant with the gravimetric sampling, distant from rainfall and irrigation events and with stable atmospheric conditions. The adimensional coefficient $\Omega$ is a constant determined on the basis of the ratio between the mass attenuation coefficient in solids and water. In this study, we adopted a value of $\Omega$ = 0.899, estimated on the basis of the specific composition of the soil of the experimental site [30]. If a soil mineralogical analysis lacks, a mean value together with its standard deviation $\Omega$ = (0.903 ± 0.011) estimated in [29] can be adopted. The $\Omega$ variability is expected to have second order effects on the estimation of gravimetric water content as the variation introduced by different $\Omega$ values would be undistinguishable within the ~0.017 kg/kg absolute uncertainty on the estimated $w$ values, which is essentially dominated by the systematic component associated to the $S^{cal}$ and $w^{Cal}$ calibration reference values [36].

While in bare soil condition gamma rays propagate only through air after having reached the soil surface, the presence of growing vegetation introduces a sizable extra attenuation that can be modelled in terms of biomass water content (BWC), i.e., an equivalent water layer which thickness varies in time as the crop evolves during its life-cycle. In order to avoid a systematic overestimation of the gravimetric water content, a correction for the BWC presence needs to be applied. The shielding effect produced by the crop system can be estimated by modeling stems, leaves and fruits as an equivalent water layer characterized by a given thickness which we express as a BWC in units of mm. The gravimetric water content corrected for the attenuation to the vegetation $w_{\gamma K}^\Lambda$ at time $t$ [36] is given by:

$$w_{\gamma K}^\Lambda(t) = \frac{S_K^{Cal} \times \Lambda_K(BWC(t))}{S_K(t)} \times \left[\Omega + w_G^{Cal}\right] - \Omega \qquad (5)$$

where the BWC correction factor $\Lambda_k$ (cps/cps) changes in time and it is given by the ratio between the $S_K$ recorded at the time t when there was a given BWC at the experimental site and the signal recorded for BWC = 0. In this study, we adopted the $\Lambda_k$ curve determined in [36] by adopting the Monte Carlo simulation method described in [29]. $\Lambda_k$ is identically equal to 1 for null vegetation cover, as expected, and is equal to 0.88 for the 9.7 mm maximum BWC estimated during the field experiment. If neglected, this missing correction would have biased the gravimetric water content by approximately a factor of 2.

The hourly average results in terms of $^{40}$K gamma count rates (in cps) and of volumetric water content $\theta_\gamma$ (in m$^3$/m$^3$) are reported in Figure 6. The latter was obtained by multiplying the gravimetric water content $w_\gamma$ for the experimental site soil bulk density (Table 1). As expected, the gamma signal and the volumetric water content are respectively negatively and positively correlated with the amount of precipitations that include both rainfall and irrigation water.

The correction for the BWC and the overall reliability of the method were tested with an ad hoc sampling campaign performed in presence of biomass and in three different soil moisture conditions (Table 2), i.e., one day before an irrigation event (24 July), one (26 July) and three days (28 July) after the same event. One set of additional validation measurements was carried out also in bare soil condition two days after a rainfall event (21 September) [36]. The gamma water contents $\theta_\gamma$, inferred from gamma signals measured in the same time interval of samples collection, were compared with the weighted average volumetric water contents $\theta_G$ inferred from the sampling campaign performed with the same scheme of the calibration process. The 0–10 cm, 10–20 cm, and 20–30 cm $\theta_G$ values were combined with weights respectively equal to 0.79, 0.16, and 0.05, determined on the basis of the gamma signal depth profile.

An average discrepancy between the experimental measurements ($\theta_G$) and estimated values ($\theta_\gamma$) of −2.1% is observed. By taking into account the statistical dispersions of the sets of gravimetric measurements, the $\theta_\gamma$ values are compatible with $\theta_G$ values at $1\sigma$ level for all the validation days. Systematic errors leading to underestimations or overestimation of the soil water content are to be excluded also in the presence of the tomato crop at the experimental site.

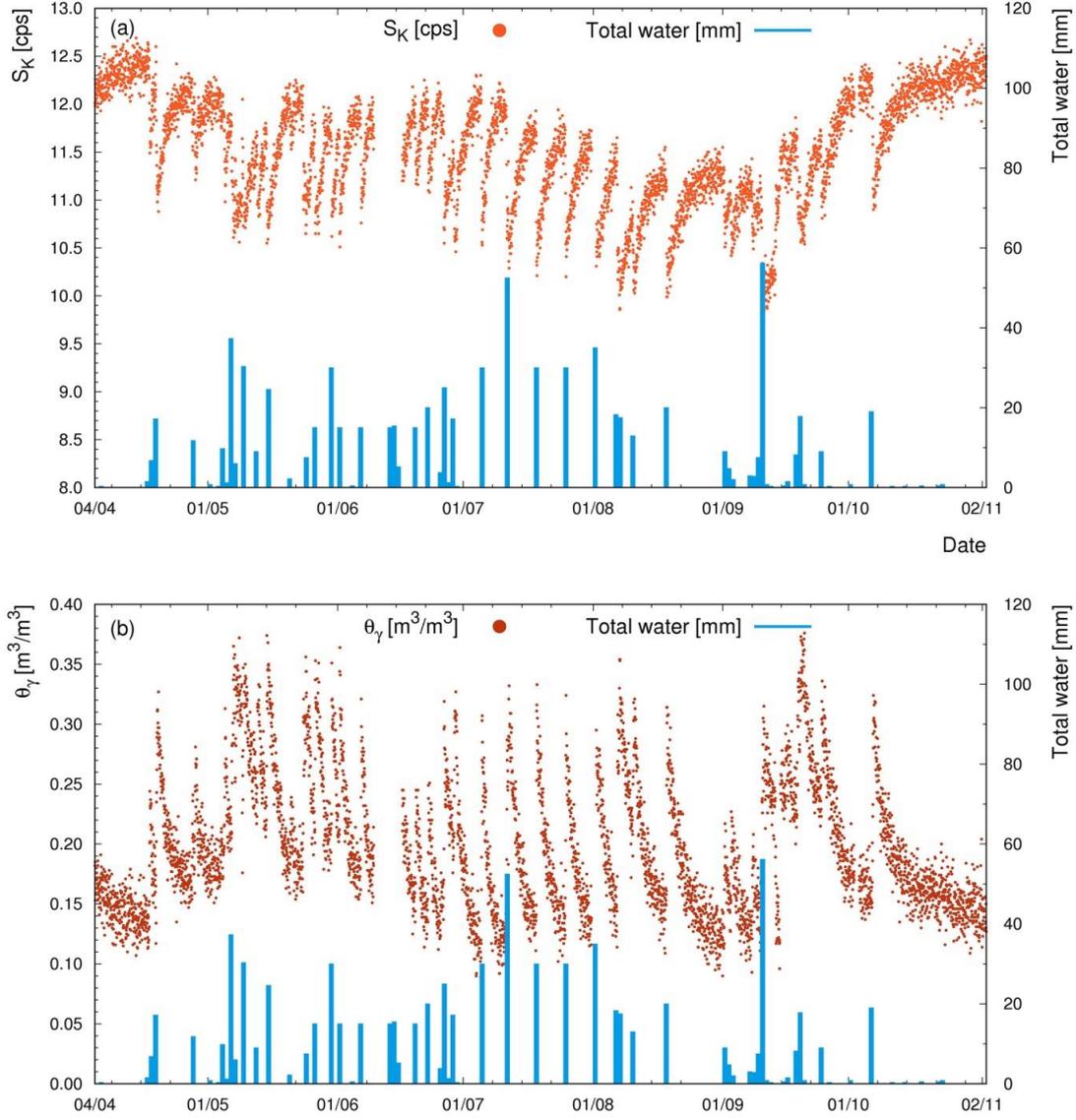

**Figure 6.** In panel (**a**) is reported the $^{40}$K gamma signal in cps ($S_K$). In panel (**b**) the volumetric water content ($\theta_\gamma$) estimated on the basis of gamma spectroscopy measurements and corrected for the attenuation due to the biomass water content. The data are hourly averaged. In both panels the daily amount of rainfall and irrigation water are reported in mm.

**Table 2.** Results of the four corroboration measurements of volumetric water content. Measurements in presence of biomass were performed one day before an irrigation event (24 July), one (26 July) and three days (28 July) after the same event. The fourth measurement (21 September) was performed in bare soil condition two days after a rainfall event. For each measurement we report the volumetric water content inferred from proximal gamma ray spectroscopy measurements ($\theta_\gamma$) together with its $1\sigma$ uncertainty, the weighted average volumetric content ($\theta_G$) determined from 16 planar sampling points homogeneously distributed within 15 m from the gamma station, the daily values simulated with the CRITeRIA ($\theta_C$) and AquaCrop ($\theta_A$) models; for IRRINET the mean daily volumetric water content ($\theta_I$) is reported only during the tomato crop season (see Section 2.3).

| Day | $\theta_G$ [m³/m³] | $\theta_\gamma$ [m³/m³] | $\theta_C$ [m³/m³] | $\theta_A$ [m³/m³] | $\theta_I$ [m³/m³] |
|---|---|---|---|---|---|
| 24 July | 0.167 ± 0.028 | 0.170 ± 0.023 | 0.134 | 0.107 | 0.154 |
| 26 July | 0.265 ± 0.028 | 0.243 ± 0.023 | 0.224 | 0.247 | 0.189 |
| 28 July | 0.189 ± 0.029 | 0.179 ± 0.023 | 0.168 | 0.152 | 0.168 |
| 21 September | 0.237 ± 0.015 | 0.245 ± 0.023 | 0.255 | 0.310 | / |

## 4. Discussions

As soil water content values provided by CRITeRIA ($\theta_C$), AquaCrop ($\theta_A$), and IRRINET ($\theta_I$) are referred to a daily frequency, we averaged the 24 hourly experimental volumetric water contents ($\theta_\gamma$) in order to obtain soil moisture dynamics curves with the same temporal resolution (Figure 7). $\theta_C$ and $\theta_A$ were determined by combining the 0–10 cm, 10–20 cm, and 20–30 cm values according to the 0.79, 0.16, and 0.05 weights described in Section 3. $\theta_I$ refers to the mean value in the variable depth horizon modelled by IRRINET (see Section 2.3.3). The four datasets show consistent temporal trends, positively correlated with the amount of precipitated water (Figure 3c).

Although the three simulated datasets never reach the saturation value $\theta_s$ (Table 1), we observed different variability ranges that reflect the different computational methods of water depth redistribution. The widest dynamic observed for the $\theta_A$ values is attributable to the tipping-bucket conceptual model implemented in AquaCrop, making the water content of the most superficial layer exceed field capacity $\theta_{FC}$ (Table 1) and reach a maximum value (0.44 m$^3$/m$^3$) close to $\theta_s$ before draining water to the deeper layers. In the summer period, when the depletion of water content due to plant water uptake becomes relevant, AquaCrop provides the lowest values (0.07 m$^3$/m$^3$), which are below the wilting point $\theta_{WP}$ (Table 1). Conversely, CRITeRIA exhibits a more homogenous depth water distribution as physically based models usually provide a more accurate computation of water fluxes across soil strata. As in CRITeRIA each soil layer is characterized by a limited water content variability (~25%), the excursion of $\theta_C$ values is confined in the (0.11–0.31) m$^3$/m$^3$ interval, almost corresponding to the ($\theta_{WP}$ − $\theta_{FC}$) range. A narrower variability (0.14–0.23 m$^3$/m$^3$) is observed in the temporal trend of IRRINET values ($\theta_I$), which are intrinsically averaged over a variable maximum depth defined as the tomato root system vertical horizon. The $\theta_\gamma$ experimental values lie in the (0.12–0.33) m$^3$/m$^3$ range, compatible with soil hydraulic properties in terms of $\theta_{WP}$ and $\theta_{FC}$ (Table 1), and are consistent with models variability almost over the entire data taking. Limited temporal misalignments between experimental and simulated daily values can be due to the averaging of gamma inferred volumetric water contents, which are sensitive to the occurring of impulsive rainfall and irrigation events (Figure 6). Conversely, simulation models cluster the output water contents with a daily resolution, making the variations unresolvable due to distinct precipitation events within a day.

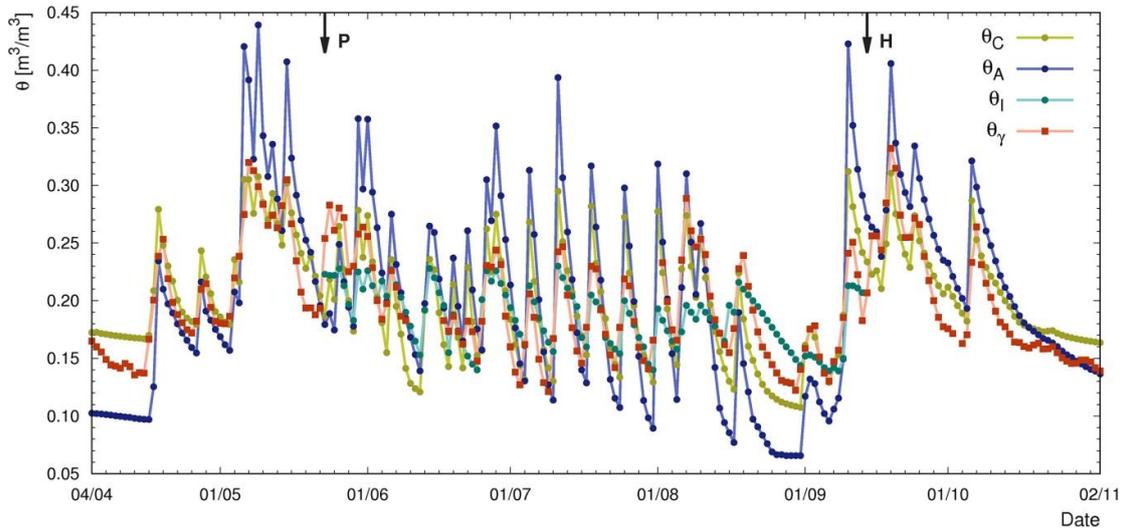

**Figure 7.** Daily volumetric water content as function of time over the entire data-taking period inferred from gamma ray spectroscopy measurements ($\theta_\gamma$), CRITeRIA ($\theta_C$) and AquaCrop ($\theta_A$), and over the tomato crop season for IRRINET ($\theta_I$). The arrows indicate the tomato crop planting (P) and harvesting (H).

The general agreement between experimental and simulated values was assessed in terms of linear regression between the datasets over the entire data-taking period and separately over the bare soil and vegetated periods (Figure 8). Together with the slope (m ± δm), intercept (q ± δq) and coefficient of determination ($r^2$), we estimated the Nash–Sutcliffe coefficient (NS) [37] (Table 3). The NS quantifies the relative magnitude of the residual variance between experimental ($\theta_\gamma$ and simulated values ($\theta_A$, $\theta_C$, and $\theta_I$) compared to the measured data variance. A good model performance is indicated by a NS value close to 1. When NS = 0 the model predictions are as accurate as the mean of the observed data, whereas for NS < 0 the observed mean is a better predictor than the model, since the residual variance is greater than the data variance. It is a useful indicator for time series as it also captures the model performance over time.

Considering the m, $r^2$, q, and NF values over the entire data taking period, CRITeRIA shows a better agreement with experimental measurements with respect to AquaCrop (Table 3). Indeed AquaCrop performed poorly, especially in presence of tomato crop when a negative NS value (−1.04) is observed and $\theta_A$ values increase on average by a factor of 2 with respect to $\theta_\gamma$ values. During the vegetated period, when also IRRINET simulated data are available, the best agreement is observed between $\theta_\gamma$ and $\theta_I$. Indeed, the goodness of fit between experimental and IRRINET simulated data is proved by a q value compatible with 0 at a 1.5σ level and an average 20% increase of soil water content with respect to experimental values. For CRITeRIA, AquaCrop, and IRRINET we obtain $r^2$ values moderately different from 1 (Table 3) which, however, do not indicate fitting models different from linear trends but highlight a relatively high dispersion of the data around the regression line (Figure 8a–e). The mean and standard deviation of percentage differences between modelled and experimental values are (1 ± 14)% for CRITeRIA, (−9 ± 35)% for AquaCrop, and (−3 ± 16)% for IRRINET. Although for all the models relatively high normalized differences respect to $\theta_\gamma$ are observed in the vegetated period (Figure 8b–f), the time series of percentage discrepancies show a scattered behavior which leads to exclude evident biases, e.g., due to the correction for BWC attenuation applied to experimental signals.

The volumetric water content values $\theta_G$ determined on the basis of gravimetric campaigns carried out in July and in September (Section 3) represent a corroboration reference for evaluating the performances of simulation models and of the gamma ray spectroscopy method. It is worth underlining that Figure 9 reports $\theta_G$ values representative of the water content of the soil at the time of the gravimetric campaign (~2 h) while $\theta_\gamma$, $\theta_A$, $\theta_C$, and $\theta_I$ refer to daily moisture levels. In the presence of tomato crops (Figure 9a), the gamma ray spectroscopy method provides the best agreement with gravimetric measurements, characterized by a 9.8% absolute mean relative discrepancy, while all the simulation models show a deviation larger than 15%. Although in bare soil condition (Figure 9b) gravimetric measurements are lower than simulated and experimental values, CRITeRIA and AquaCrop show coherent temporal trends, enclosing the time series inferred from gamma ray measurements.

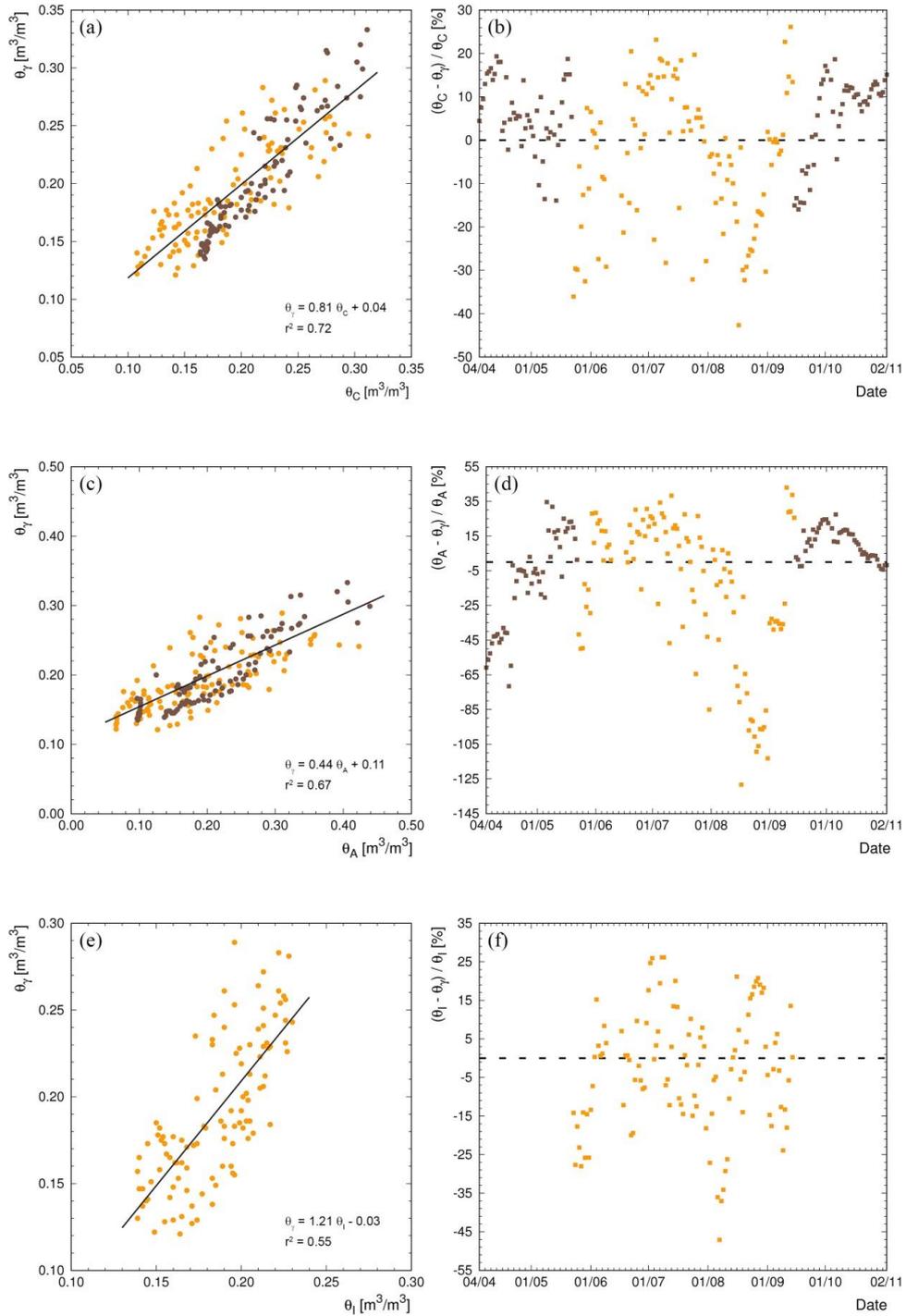

**Figure 8.** Scatter plot of the daily volumetric water contents inferred from gamma ray spectroscopy ($\theta_\gamma$) versus the daily volumetric water content estimated with CRITeRIA ($\theta_C$) (panel **a**), AquaCrop ($\theta_A$) (panel **c**) and IRRINET ($\theta_I$) (panel **e**). For each scatter plot the regression line is superimposed and its corresponding equation is reported. The percentage differences with respect to $\theta_\gamma$ values are shown for $\theta_C$ (panel **b**), $\theta_A$ (panel **d**), and $\theta_I$ (panel **f**). Different colors are assigned to data points in bare soil condition (brown) and in presence of the tomato crop (orange).

**Table 3.** Parameters of the linear regressions between the daily volumetric water content inferred from gamma ray spectroscopy measurements and from simulation models. Together with the slope (m ± δm), intercept (q ± δq) and coefficient of determination ($r^2$), the Nash–Sutcliffe efficiency (NS) is reported for three different datasets: in presence of the tomato crop, in bare soil condition, and for the whole data-taking period.

|  |  | CRITeRIA | AquaCrop | IRRINET |
|---|---|---|---|---|
| In presence of the tomato crop | m ± δm | 0.68 ± 0.05 | 0.37 ± 0.03 | 1.21 ± 0.11 |
|  | q ± δq [m³/m³] | 0.06 ± 0.01 | 0.12 ± 0.01 | −0.03 ± 0.02 |
|  | $r^2$ | 0.65 | 0.53 | 0.55 |
|  | NS | 0.51 | -1.04 | 0.51 |
| Bare soil condition | m ± δm | 1.15 ± 0.04 | 0.57 ± 0.03 | / |
|  | q ± δq [m³/m³] | -0.04 ± 0.01 | 0.08 ± 0.01 | / |
|  | $r^2$ | 0.87 | 0.80 | / |
|  | NS | 0.81 | 0.27 | / |
| Whole period | m ± δm | 0.81 ± 0.04 | 0.44 ± 0.02 | / |
|  | q ± δq [m³/m³] | 0.04 ± 0.01 | 0.11 ± 0.01 | / |
|  | $r^2$ | 0.72 | 0.67 | / |
|  | NS | 0.69 | −0.27 | / |

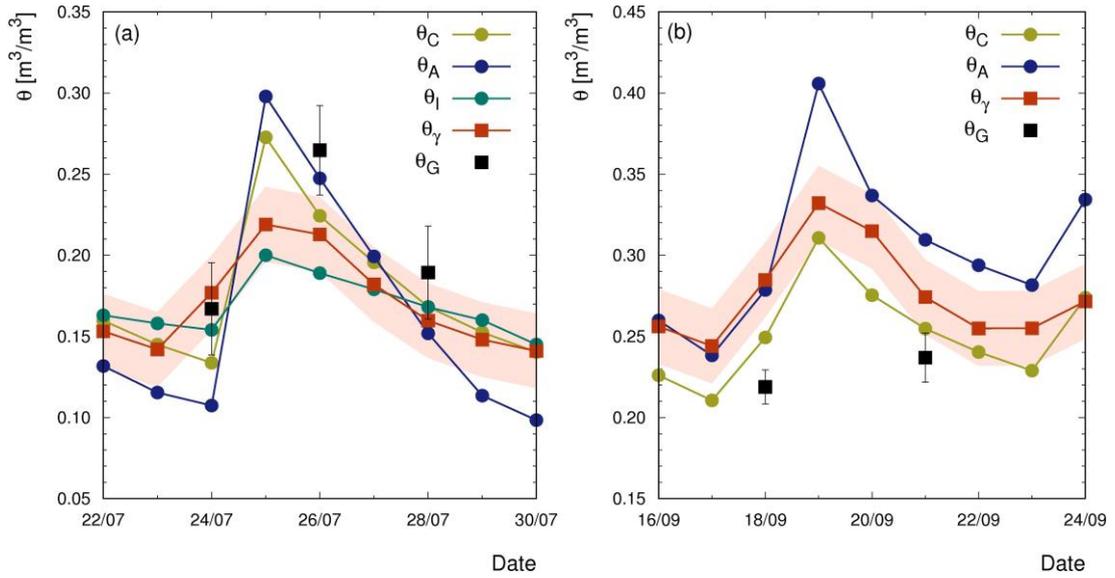

**Figure 9.** Weighted average volumetric water contents derived from the gravimetric campaigns ($\theta_G$) carried out in the late morning (10 a.m.–12 a.m.) during the vegetated period (panel **a**) and in bare soil condition (panel **b**). The $\theta_G$ error bars correspond to the weighted average standard deviation. Daily volumetric water contents ($\theta_\gamma$) are reported together with their 1σ uncertainty. Daily simulated values with the CRITeRIA ($\theta_C$), AquaCrop ($\theta_A$), and IRRINET ($\theta_I$) models are also plotted.

## 5. Conclusions

In the perspective of optimizing agricultural management with a sustainable use of water, one of the current challenges is performing continuous and reliable soil moisture monitoring at field scale with non-destructive and real time techniques. In this context, we performed a proximal gamma ray spectroscopy experiment by installing a permanent station in a tomato field which allowed for assessing soil water content dynamics over a seven-month period. The main results of this work are as follows:

(i) Proximal gamma ray spectroscopy is an excellent method for a non-stop tracing of soil water content at an intermediate spatial scale between punctual and satellite fields of view;
(ii) Once a reliable calibration is provided through direct measurements, soil water contents inferred from gamma ray spectroscopy do not require detailed soil and crop parameterization and are characterized by relatively low uncertainties;
(iii) While soil–crop system models simulate soil dynamics with a daily resolution, the proposed method is able to provide reliable higher frequency estimations sensitive to transient soil moisture levels, as proved by the excellent agreement with direct gravimetric measurements;
(iv) Proximal gamma ray spectroscopy gives a satisfactory description of soil water content over time also when compared to simulation data, showing that the combination of accurate soil water content measurements and water budget computation with crop models can be effective tools for water resources and irrigation planning.

**Acknowledgments:** This work was partially founded by the National Institute of Nuclear Physics (INFN) through the ITALian RADioactivity project (ITALRAD) and by the Theoretical Astroparticle Physics (TAsP) research network. The authors would like to acknowledge the support of the Project Agroalimentare Idrointelligente CUP D92I16000030009, of the Geological and Seismic Survey of the Umbria Region (UMBRIARAD), of the University of Ferrara (Fondo di Ateneo per la Ricerca scientifica FAR 2016), and of the MIUR (Ministero dell'Istruzione, dell'Università e della Ricerca) under MIUR-PRIN-2012 project. The authors thank the staff of GeoExplorer Impresa Sociale s.r.l. for their support and Renzo Valloni, Enrico Calore, Fabio Sebastiano Schifano, Claudio Pagotto, Ivan Callegari, Gerti Xhixha, and Merita Kaçeli Xhixha for their collaboration which made possible the realization of this study. The authors show their gratitude to Giovanni Fiorentini, Ilaria Marzola, Michele Montuschi, and Barbara Ricci for useful comments and discussions. We would like to thank Carmine Gerardo Gragnano for particle size distribution analysis. The authors thank the Università degli Studi di Ferrara and INFN-Ferrara for the access to the COKA GPU cluster.

**Author Contributions:** V.S., M.A., M.Ba., and F.M. conceived and designed the work as it is. V.S., M.A., S.A., M.Ba., M.Bi., C.B., E.C., K.G.C.R., D.S., and F.M. realized the experimental setup and performed the measurements with the support of B.F. and V.G. The software for processing gamma data and the spectroscopy analysis was carried out and tuned by M.Ba. and E.C.; the simulations with soil–crop system models were performed by M.Bi., F.T., and G.V. (for CRITeRIA); by M.A. and K.G.C.R. (for AquaCrop); and by S.A. and D.S. (for IRRINET). V.S., M.Ba., and F.M. took the lead in designing and composing the manuscript with the input from all the authors. All the authors critically revised and provided final approval of the version submitted.